# Spatially Resolved Evaporative Patterns from Water

**Federico Ienna, Hyok Yoo, and Gerald H. Pollack**
Department of Bioengineering, Box 355061, University of Washington, Seattle, Washington 98195, United States

## Abstract

Unexpectedly distinct patterns in evaporation were observed over heated water. Although the patterns had chaotic aspects, they often showed geometric patterns. These patterns bore strong resemblance to the infrared emission patterns observable with a mid-infrared camera focused on the water surface. This similarity puts constraints on the mechanism of evaporation, and leads to a general hypothesis as to the nature of the evaporative process.

## Introduction

Water evaporation plays a critical role in numerous large-scale atmospheric, oceanic, and environmental phenomena as well as in small-scale interfacial and colloidal phenomena. Despite evaporation's broad significance, a century of investigation has failed to yield a full description of the underlying process.

The currently accepted molecular model rests on concepts offered by Hertz and Knudsen, based on classical kinetic theory of gasses (1-3). While this approach was initially successful for explaining experimental observations on several viscous liquids, it has yet to prove useful for water. In part, this is because experimental values of the evaporation coefficient have varied widely. Numerous studies through the mid-1980s produced values ranging over two orders of magnitude (see [4,5] for summary). Further attempts to measure this coefficient have dwindled, leaving the true value uncertain.

A potential difficulty in obtaining reliable measures of evaporation coefficient may arise from the chaotic nature of Marangoni-Benard convection occurring at the air-water interface. Although well-known studies by Cammenga and others have shown no evidence of surface tension-driven convection in water (6, 7), more recent studies have shown otherwise (8-10). These latter studies seem in accord with recent infrared imaging data, where, at least in water droplets at elevated temperatures, spatially resolved convection cells are clearly observed (11).

In this report we show, through a simple experimental approach. that Marangoni-Benard cells are clearly seen in water, and that they play a definable role in evaporation.

## Methods

Three different methods were used to visualize evaporation, condensation, and heat-flux patterns above a pre-heated water bulk.  The three methods used were (*a*) forward-looking infrared visualization; (*b*) forward scattering by floating droplets; and (*c*) laser-plane forward scattering by airborne condensation particles.  Each method is detailed below, and illustrated in Figure 1.  In some cases experiments were conducted using multiple lighting and recording configurations simultaneously.

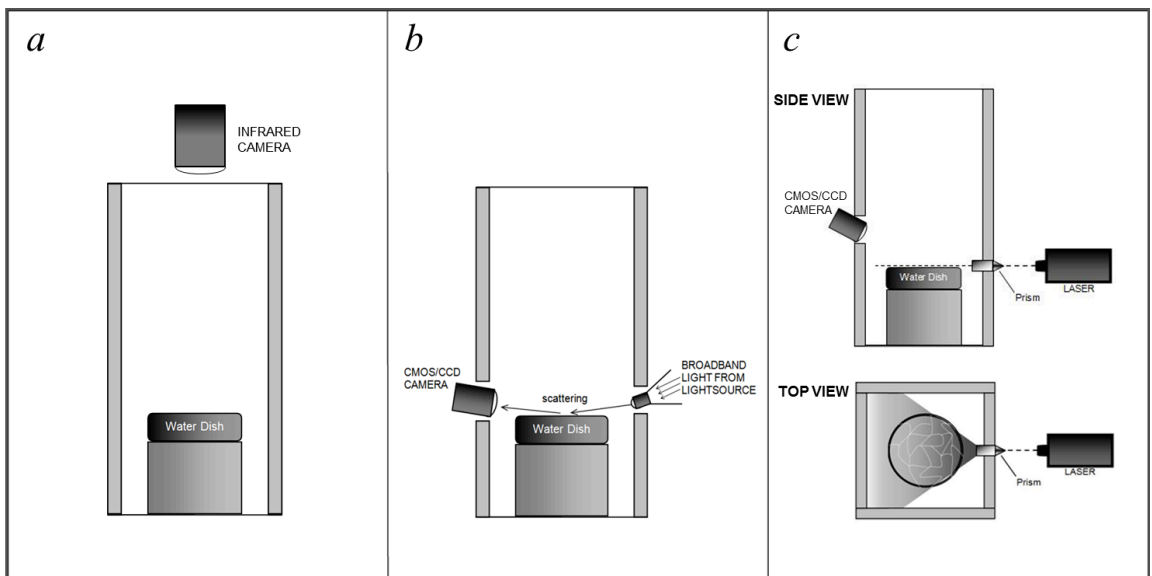

**Figure 1:** Experimental setups: (*a*) forward-looking infrared visualization; (*b*) forward scattering by floating droplets; and (*c*) laser-plane forward scattering by airborne condensation particles (side and top view).

**Preparation of sample**

All experiments were conducted using a shallow Pyrex Petri dish (10-cm diameter X 1-cm height).  The Petri dish was coated with black paint on its outer walls in order to increase contrast and improve picture quality.  The dish was filled with pre-heated, distilled, de-ionized water (Type 1, HPLC grade, 18.2 MΩ) obtained from a standard water purification system (Diamond TII, Barnstead) prior to pre-heating.  Pre-heating of the water was performed in a separate Pyrex flask (No. 70000, Pyrex Vista) placed onto a standard hotplate heater (Isotemp, Fisher Scientific), with the top of the flask covered to reduce water loss due to evaporation.

The water was heated until a temperature of 100 °C was reached, at which point it was poured slowly into the Petri dish in order to minimize air intake from splashing or bubbles. The volume of water used was 154-ml, which is the volume of heated water required to fill the dish to the brim while maintaining a minimally curved meniscus at the top.

During experimentation, the water in the dish was allowed to naturally cool toward room temperature as various observations and recordings were performed. Protocols for sample preparation, experimentation, and recording were essentially identical for all experiments, except for lighting and recording devices.

**Forward-looking infrared visualization**

The infrared radiation emitted from the water surface was recorded using an infrared sensing camera located directly above a pre-heated water container. The camera used was a ThermoVision SC-6000 (FLIR Systems) with a spectral sensitivity range between 3-µm and 5-µm in wavelength. The temperature sensitivity was 50 mK. The camera was placed approximately 40 cm above the pre-heated water, looking directly downward and focused on the top surface of the water as it cooled naturally. The software used for visualization of the data was ExaminIR 1.2.0.1076 (FLIR Systems). ExaminIR generates a false-color image, with different colors or shades representing the intensities of the infrared radiation emitted by the subject and received by the InSb detector. In those cases where the infrared camera was used, it was the main method for determining the average temperature of the water surface.

**Forward scattering by floating droplets**

Surface droplets were visualized by taking advantage of their natural light-scattering properties. These droplets were present on the water surface between surface temperatures of approximately, 95 °C and 50 °C, and persisted as small (about 7-µm diameter) spheres. Thus, they are excellent scatterers in the Mie-scattering realm, scattering strongly in the forward direction.

To accurately observe the droplets' distribution, size, and other properties, a simple broadband light-source (Model I-150, Cuda) was positioned on one side of the water container so as to shine intensely onto the surface from a shallow angle, and a standard CMOS digital camera (Nikon D300s) was placed directly opposite to the light source, thus maximizing the light scattered off of the droplets and into the camera lens. The latter was a Nikon Nikkor 35-mm f/2.0 AF-D SLR camera lens. This simple set-up provided a reliable and consistent method for observation and measurement of droplet features.

For measurements requiring higher magnification, a microscope was used in lieu of the camera lens. In those instances, images were recorded with a smaller CCD sensor monochrome digital video USB camera (EO-0413M MONO USB, Edmund Optics), allowing for higher frame-rate recordings.

**Laser-plane forward scattering by airborne condensation particles**

A flattened laser plane (or "light-slice") was used for observation of airborne particles above the water surface. Use of the light-slice allowed for the scanning of thin sections of the particle cloud at selected heights above the water bulk by forward scattering. A green laser-diode (<5-mW output, 532 ± 10-nm) was passed though a laser line generator (75° full fan angle, Edmund Optics), which flattened the beam into a thin horizontal plane. The effective thickness of the laser plane above the water dish was 1.0 ± 0.25-mm, and the plane was fixed at a height of 0.75 ± 0.25-mm. Images were recorded with the same monochrome CCD digital video USB camera used in other forward scattering experiments.

The height of the laser-illuminated plane above the water surface was carefully measured before each experiment. Several heights were tested, but the height used most frequently was 1.0 ± 0.25-mm. This is the smallest vertical distance that could be consistently attained without causing the laser to come in contact with the water surface. Patterns in the particles' distribution could thus be resolved spatially and temporally using this set-up.

**Ultra-high-speed recording of droplets**

Surface droplets were at times observed to depart from the water surface by either lifting off and becoming airborne, or by seemingly disappearing. In the latter cases, it was clear that the recording frame-rate was not high enough to discern what was happening. Therefore, ultra-high-speed recordings were obtained at frame-rates ranging from 20,000 to 75,000 fps using a high-speed CCD video camera (Fastcam APX-RS, Photron).

The experimental set-up was identical to the one used for all other recordings of surface floating droplets, the only difference being the ultra-high-speed camera. While the camera was capable of reaching recording speeds up to 200,000 fps, the limiting factors were resolution and lighting requirements. Thus, 75,000 fps was the highest practical recording rate that could yield successful imaging.

# Results

**Surface Visualization by Infrared Imaging**

Direct recordings of infrared emission from the water surface have yielded the well-known patterns often referred to as Marangoni-Benard cells. An example is shown in Fig. 2. These mosaic-like cells are most evident at temperatures above approximately 30° C. Infrared emission is high *within* the cells, while it is lower at the darker borders. The insides of the cells can thus be said to be warmer than their boundary.

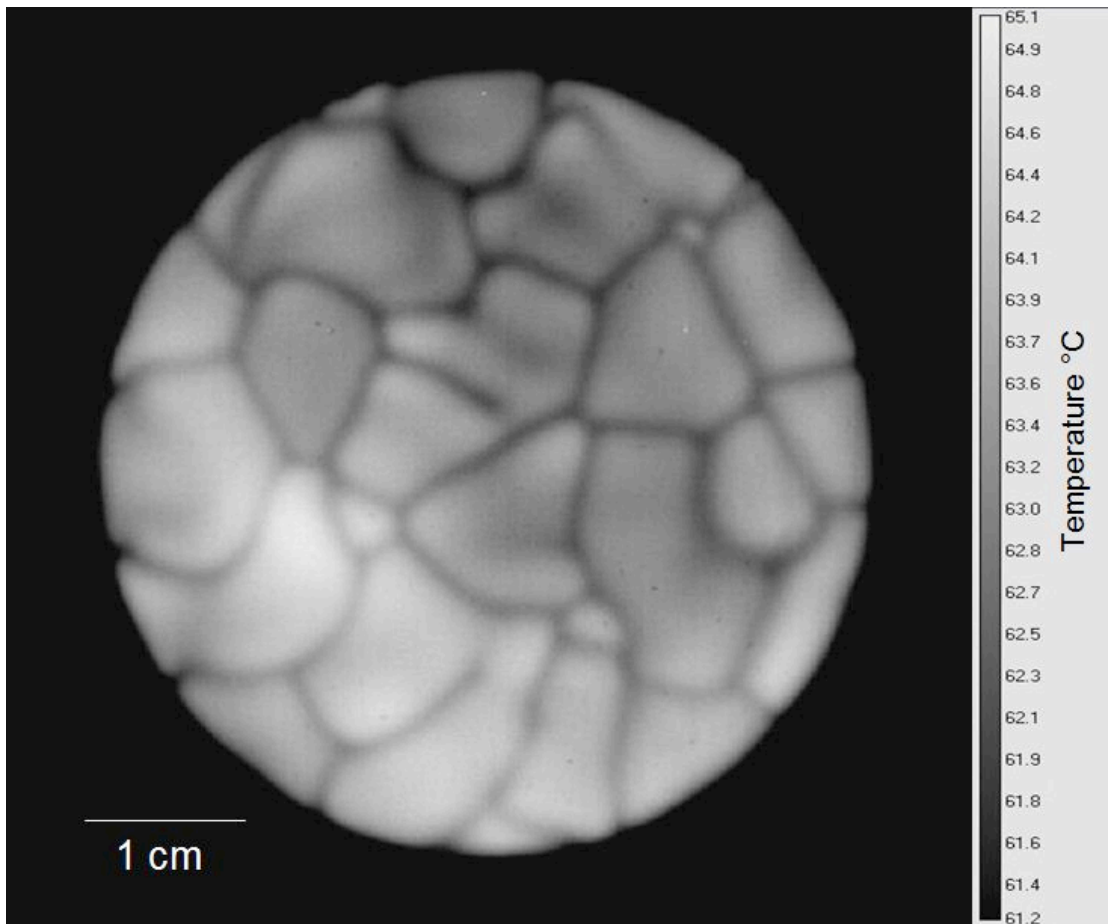

**Figure 2**: Infrared image of the surface of water, viewed from directly above. Corresponding temperatures are shown on the right.

Images obtained during the course of natural cooling are shown in Figure 3. At the highest temperatures, the patterns are most distinct but least stable, mixing and moving about rapidly. At lower temperature some mixing still occurs, including the joining and dividing of individual cells. Mixing and shifting occurred at lower temperatures as well, although less frequently. Nearest to room temperature, any mixing and shifting was rare.

The cells and their boundaries also became less apparent at the lower temperatures. As the surface approached room temperature, contrast between the light cells and their dark boundaries became progressively weaker, the lines sometimes diminishing beyond the camera's resolution at approximately 22°C.

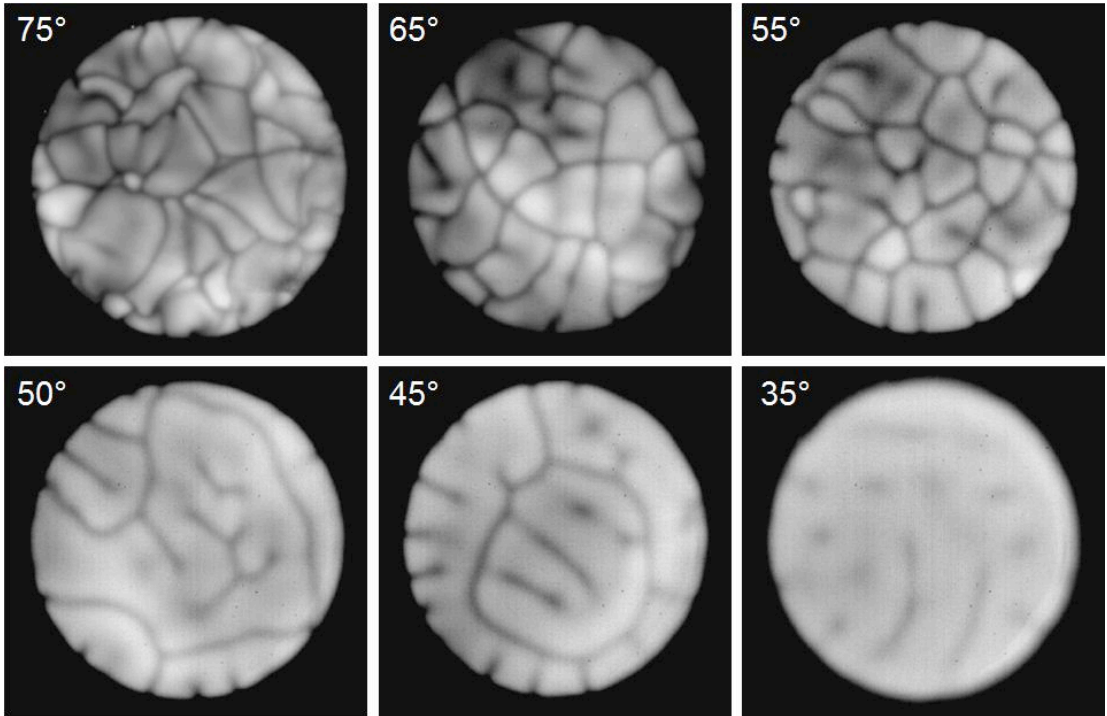

**Figure 3:** Temperature dependence of surface images. The median surface temperature for each snapshot is noted in each frame in degrees Celsius. As the water cools, the cells stabilize, diminish in number, and eventually become less evident.

The patterns of Figures 2 and 3 do not exist at the air-water interface alone. The gallery of oblique and side views shown in Figure 4 demonstrate that the dark boundary lines lying in the detectable range of the camera can be seen to run down into the vessel. Some lines run almost vertically, while others are angled. Sometimes the lines may fuse. Videos taken over tens of seconds show that the lines bend, undulate, disappear, reappear, and even occasionally turn horizontal. Thus, the lines not only have vertical extent but also are quite dynamic.

**Surface Visualization by Visible Light Scattering**

When a high-intensity light source was placed at a shallow angle to the water surface, a mosaic pattern virtually identical to the one seen with the infrared camera was visible (Fig. 5). This appeared to be the result of forward scattering. Small droplets persisting on the water surface provided an organized array of visible scatterers. The droplets appeared to reside more-or-less within the boundaries of the Marangoni-Benard pattern, making the pattern easily observable to the naked eye.

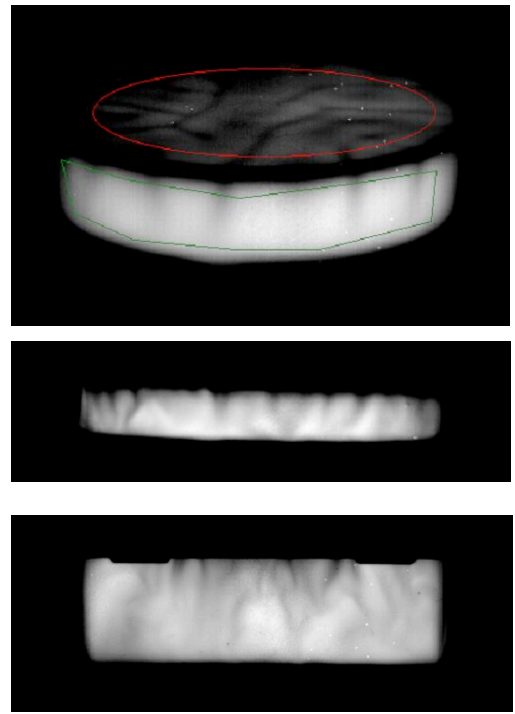

**Figure 4**. Oblique and side views obtained with the infrared camera.

The images of Figure 5 were taken simultaneously. Hence, they represent a side-by-side comparison of the mosaic patterns and the floating vesicle distribution pattern. The cameras' viewing angle was different for the two experiments—the infrared camera was located directly above the water surface, while the visible-light camera was located at a shallow angle to the surface. To provide a good side-by-side comparison, it was therefore necessary to vertically stretch the visible light image (Figure 5*b*) to yield the same aspect ratio as the infrared image. Clearly, the two patterns match closely.

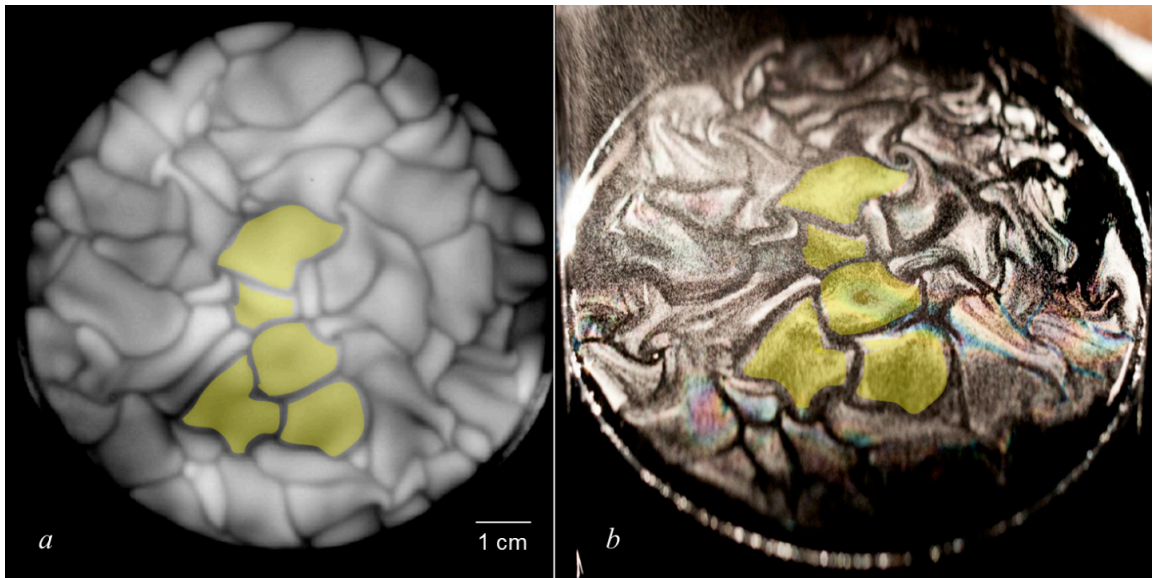

**Figure 5:** Side-by-side comparison between the emitted infrared mosaic pattern (a) and the visible distribution of floating droplets (b). Several of the cells in both images are highlighted in yellow to facilitate the similarity.

The floating droplets were seen when the water temperature was between approximately 50 °C and 95 °C. However, it was only in the higher range, roughly between 65 - 95 °C, that they organized into the recognizable patterns. In this temperature range, the droplets were lodged mostly within the cell areas, with fewer along the cell boundary areas. Below 65 °C, the droplets were fewer in number with no recognizable pattern, while below ~50 °C and with the magnification used, no droplets were observed at all.

Representative close-up images of droplets at a series of magnifications are shown in Figure 6. In this particular case dark zones surround the droplet cluster on either side; hence, the droplets appear to reside within a Marangoni-Benard cell. From microscopic observation, the droplets appear to be liquid water droplets, much like the airborne aerosol droplets that form visible clouds above a hot water bulk.

The shape of the floating droplets—appears spherical (see figure 6*c* inset). All observed droplets maintained this shape throughout all experiments. To measure their size, more than 300 droplets were photographed individually in standard conditions through a 5X objective, and their diameters were measured using the software ImageJ (NIH). Mean diameter at 80 °C was 6.51 +/- 1.57 µm (SD); *N* = 319.

The sizes were also plotted against the mean water temperature (Figure 7). Fitted linearly, the trend indicates a modest decline in droplet size as temperature diminishes. On the other hand, the standard deviations are too high to be certain of a true functional dependence. Interestingly, the standard deviations were smaller in the lower temperature range, implying a more deterministic process in that temperature range.

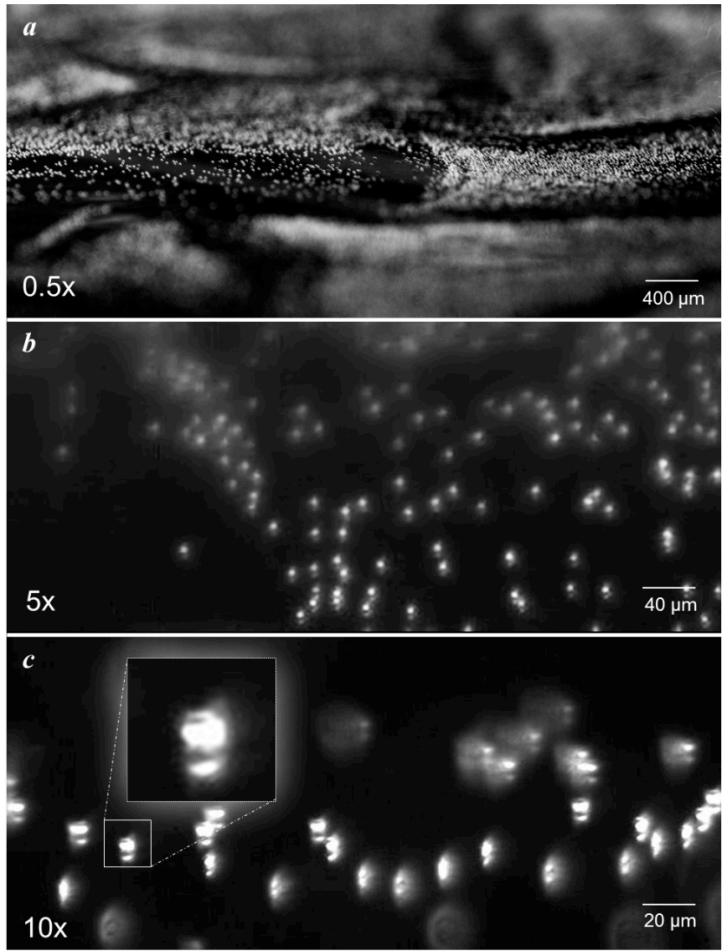

**Figure 6:** Three close up images of the floating droplets photographed using 0.5x, 5x, and 10x objectives (*a, b,* and *c,* respectively). The inset in panel *c* shows a closer view of a well-focused droplet and its reflection below it.

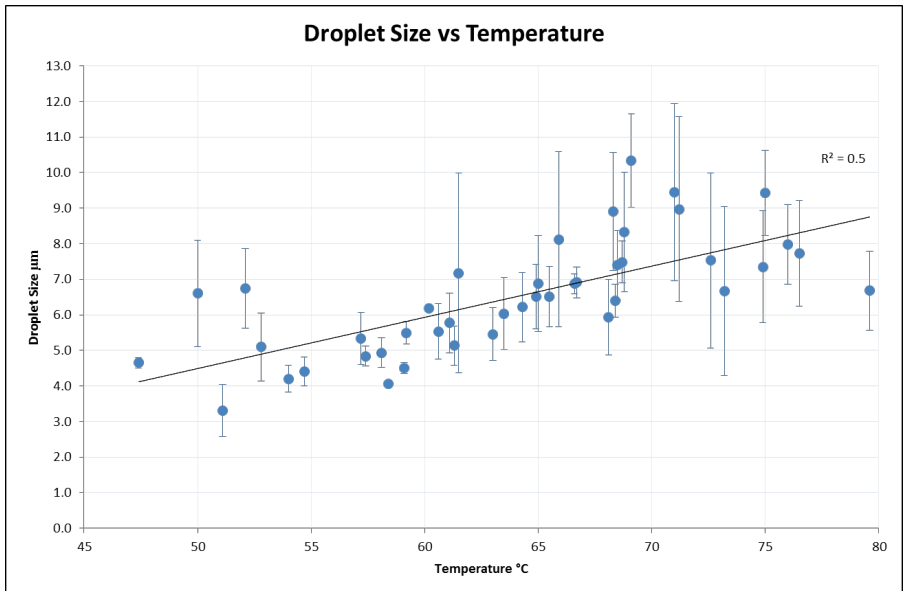

**Figure 7:** Droplet size (diameter) vs bulk water temperature. Each point corresponds to the average diameter of all droplets measured at the respective temperature, ranging from 2 to 15 droplets for each point. The vertical bars are the standard deviations for each point.

**Evaporative Patterns**

The vapor pattern was recorded as close to the warm liquid surface as technically feasible. Representative records are shown in Figure 8. The patterns are not amorphous, as conventionally expected. They frequently showed distinct patterns such as those in the figure. The bright regions correspond to the vapor aerosol, whose particles scatter light and create the image.

Most generally, the vapor patterns consisted of contiguous loops, much like the patterns seen with the infrared camera. Line thicknesses were generally much narrower than loop diameter. Sometimes the lines curved, but straight segments were also seen, as in panel (b). Often three lines would merge at one point. Sometimes the lines would terminate without completing the loop. These mosaic-like features of the vapor were characteristic also of the water-surface mosaic seen with the infrared camera.

When the camera angle was changed, we could often detect the individual aerosol droplets that made up the vapor loops.

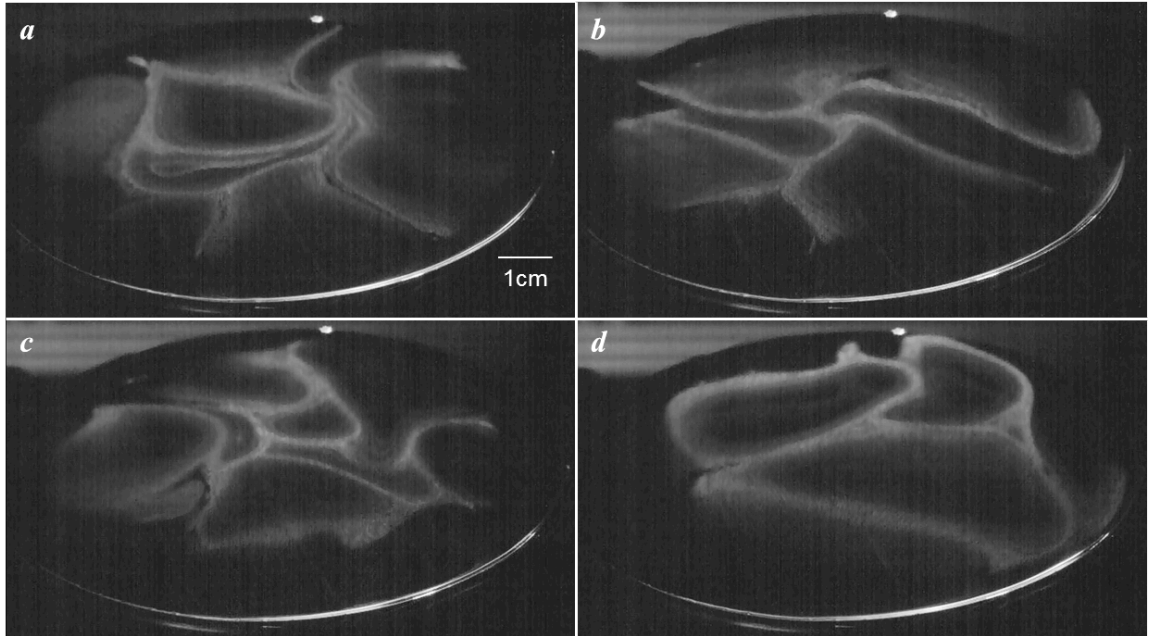

**Figure 8:** Examples of airborne aerosol patterns imaged by the forward scattering of a planar laser beam. The white lines represent a high concentration of airborne particles scattering the laser light. Regions where there are no airborne particles to scatter the light are seen as dark.

Not all vapor images were as distinct as those shown in Figure 8. At temperatures very close to the boiling point, where the dynamics were dominated by heavy and chaotic evaporation, the patterns were often hardly visible. At lower temperatures the distinctive patterns seemed to come and go, being clear as in Figure 8 for some seconds and then disappearing for others.

A pattern's persistence time was also variable. The pattern might persist stably for as long as several seconds or for as little as half a second as the vapor rose. Generally, persistence was slightly longer at lower temperatures, but persistence-time variability was high at all temperatures.

The temperature dependence of the vapor features was much like that of the water-surface features. At surface temperatures near boiling, the vapor line patterns moved vigorously and chaotically and as a result it proved difficult to record them with the available apparatus. As the surface temperature diminished to 50° - 60°, the vapor patterns became more stable, and hence more easily recordable. As the bath cooled still further (45 – 50 °C), the patterns became even more distinct and stable, retaining their shape for longer. With further temperature diminution and decreased evaporation rate, the evaporative patterns also become less and less visible. These behaviors are similar to the respective behaviors of the water surface patterns.

**Droplet Lifting and the "Ripping-line" Effect**

All experiments showed droplets either already residing on the surface, or settling onto the surface from above. Droplets could be seen to descend from an airborne location above the surface. Settling presumably came from downdrafts, replacing the upward moving air and vapor. The droplets seemed to descend and settle in grouped clusters of 5-20 rather than one at a time. Thus, clustering may impart some feature that promotes settling.

Once the droplets settled, they persisted on the surface for many seconds, sometimes minutes. They appeared to float directly on the surface. When finally they disappeared, they did so by one of two processes: by coalescing with the water beneath; or by lifting off the surface and becoming airborne. Coalescence itself was extremely fast, and was resolvable only with 75,000 frames/second recording rate. The lift-off process, on the other hand, was much slower and was resolvable by standard techniques.

The lifting process was closely related to the surface mosaic pattern. It occurred preferentially at a boundary, as opposed to within the interior of a Marangoni-Benard cell. Once it occurred, it left behind a distinct pathway of droplet-free space on the surface, *exactly* corresponding to a mosaic line. This was confirmed by comparing the visible scatter pattern and the infrared radiation pattern simultaneously during a lift-off event.

In the vast majority of cases, lifting took place along a line, starting on one end and ending at the other; it was as though the droplets were being "ripped" from beneath with a piece of scotch tape, hence the term "ripping-line" effect. The upward moving array of droplets could be seen to ascend from this line like an upward waterfall.

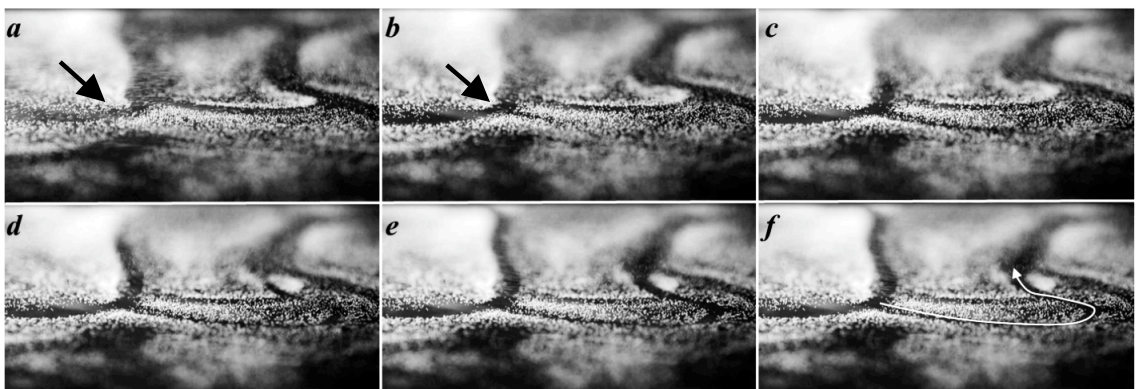

**Figure 9:** Time-series demonstrating the ripping line effect. Each panel corresponds to a 0.125 second increment of time. The arrows in panels *a* and *b* point to where a newly forming clear line is being generated. The line to the left of the arrowhead is already darker in *b*, compared to *a*. The line continues to the right in panel *b*, and continues to extend to the right in panels *c*, *d*, and *e*.  Panel *f* shows a trace of the completed path and the newly formed line.

Figure 9 shows a time-series of this effect.  We can see a new dark line, just beginning to form where the arrow is pointing in panels *a* and *b*.  The new line starts to become more distinctly visible near the arrow in panel *b*, and builds towards the right through panels *c* and *d*, eventually curving upward in panel *e* and reconnecting to a previously existing line.  In panel *f* we trace the shape of the newly formed line.  The arrow in panel *f* points to the direction that the line continued to follow.

It is important to note that the line being formed by the lifting droplets was *already in existence* in the corresponding infrared images before the lift-off event.  Thus, it is not a completely new mosaic line that is formed, but rather it is the droplets that leave the surface containing a pre-existing line underneath. Following liftoff, the line in the visual pattern becomes darker and more distinct.

## Discussion

The results demonstrate a linkage between spatial patterns detectable in warm water and the evaporative patterns emerging from that water. Several aspects of the observations are unexpected.

The first unexpected feature is that the evaporative patterns appear as rings. Generally, the images showed several contiguous rings, whose boundaries were either curved or straight. These patterns were imaged as close to the liquid surface as was technically feasible; hence, the rings appeared to rise directly from the water surface, although the visual observations alone could not exclude the possibility that they formed in the space just above the surface.

The rings had considerable vertical extent. A succession of thin planar slices recorded as the vapor rose showed consistent features, although subtle shifts occurred progressively with each frame. The consistency persisted for durations ranging from a half second to several seconds. Since each frame captures newly rising vapor, this consistency implies that each "ring" is in fact a section of a tube with considerable vertical extent.

These tubes may represent the vapor clouds that one can ordinarily see rising from a cup of hot liquid. These puffy clouds often emerge one at a time in rapid succession. They are commonly of centimeter size, and often show empty regions in their centers.

The second notable feature of the results is the vapor pattern's similarity to the surface pattern, the latter commonly known as Marangoni-Benard cells (12). Both patterns are mosaics, and both showed similar features. Most often the surface patterns are detected with infrared cameras or fish skin flakes, but here we have shown that the patterns are detectable also by visible light imaging at low incidence angle (Fig. 5).

Thus, the rising vapor rings appear to originate from the surface rings. This supposition was reinforced by imaging the vesicles rising from the rings. (We adopt the generic term, "vesicles" because the nature of the spherical entities rising from the liquid is not yet established.) The vesicles appeared to "rip off" existing mosaic lines in vertical sheet-like fashion, leaving the lines devoid of vesicles. The simplest interpretation is that those rising vesicles become the vapor. This would explain the vapor's ring-like shape, and would explain why the vapor rings are built of aerosol droplets (as are clouds).

The only serious differences discerned between the surface patterns and the vapor patterns were the relative size of the cells and their dynamics. The vapor lines were longer than the surface-mosaic lines; hence the vapor cell areas were larger. One reason might be that, at least in liquid, cooling increases cell size (see Figure 3). Thus, as the tube rises into cooler air, it might expand. A 20° to 30° temperature drop, according to Figure 3, could easily produce a doubling of cell size. Wind shear and air convection might play a role as well: Since the rising vapor-containing hot air must be replaced by cooler air descending from above, air convection is inevitable. The resulting circulation will exert lateral forces on the vapor, inducing distortion and curvature (compare Figures 3 and 8); this will tend to lengthen the lines, especially given the relatively low air viscosity. The lateral force will also cause the vapor to be more dynamic than the surface mosaics, a feature that was observed.

In other words, the size differences between vapor and surface cells seem potentially explainable by natural processes, although further study of the differences are in order.

**Structure of the Vapor**

From the discussion above, it appears that droplets rising from the surface-mosaic lines may form the visible vapor. The surface droplets were commonly 5 - 10 µm in diameter; thus, the rising vapor consisted of droplets/vesicles on the order of several micrometers. Considerably larger than the wavelength of incident light, those vesicles would scatter considerable light, thereby accounting for the relatively distinct appearance of the vapor images.

A point of relevance is that the detection method is unable to detect any smaller vapor constituents that do not scatter appreciably. Thus, single molecules or sub-micron aerosol particles might have augmented the visible vapor patterns, but escaped detection. No information on any such putative vapor components could be made.

A question is whether the ripping surface droplets were of sufficient quantity to account for the full vertical extent of the vapor tube. Surface droplets were limited in number

(Fig. 6). The rising tubes, by contrast, were sufficiently long — they were seen in numerous successive fames — that numerous droplets would be required to build their full length. Hence, if the tubes comprise aerosol droplets, then extra droplets must come from somewhere to build the full structure.

One possibility is that such droplets could have condensed from random locations in the air immediately above the water surface. This would require a mechanism for drawing those aerosol vesicles toward the visible ring. The horizontal velocity would need to be high, since the ring is detectable closely above the water surface. The nature of the force underlying any such rapid horizontal displacement is not obvious; nor is it obvious why those droplets remained undetectable in the video images; mainly, the droplets came directly from the lines.

An alternative possibility is that the needed vesicles come from deeper into the liquid. The mosaic lines did not reside on the surface alone; images such as that of Figure 4 imply that they extend deeper into the liquid. Therefore, the vesicles might likewise have come from deeper into the liquid. Those deeper vesicles would explain why the droplet sheets rising from the surface lines persisted for as long as they did, and perhaps also why the vertical lines in the liquid were unexpectedly dynamic. Clearly, this is a hypothesis that requires further study.

Subsequent studies will need to address a number of issues. One of them is a more detailed and quantitative relation between water and vapor dynamics. Another is the exact nature of the water-mosaic tubes and their dynamic nature. Still another issue is why the vapor vesicles do not immediately disperse as they rise into the air. Fuller understanding of these issues will help unravel details of how the water transitions into vapor.

**Conclusions**

Two main conclusions can be drawn from the data. The first is that the vapor just above the evaporating surface has distinctive shape; it is not amorphous. In the context of conventional expectations, such well-defined geometric shapes seem surprising.

The second conclusion is that the vapor patterns resemble the mosaic patterns in the water beneath. Since the vapor rises from the water, the water-surface patterns evidently give rise to the vapor patterns, and this was demonstrated by the direct observation of surface-droplet "lift-off" to form the vapor pattern. Thus, evaporation appears to be a more deterministic process than had previously been thought.